\documentclass[letterpaper, 12 pt]{article}
\usepackage[a4paper, total={6in, 8in}]{geometry}
\usepackage{graphicx} 
\usepackage{amsmath}
\usepackage{graphicx}
\usepackage{cancel}
\usepackage{lipsum}
\usepackage{cite}
\usepackage{dirtytalk}

\title{Geodesic congruences in acoustic spacetimes and the role of Raychaudhuri equation}
\author{Akshat Pandey\footnote{apandey.physics@gmail.com} \\ \normalsize Department of Physics, Shiv Nadar Institution of Eminence \\ \normalsize Greater Noida, Uttar Pradesh-201314, India}
\date{}
\begin{document}

\maketitle

\begin{abstract}
It has been known that the propagation of sound in fluids can be used to model acoustic spacetimes. These acoustic spacetimes offer analogue models for gravity. We use the Raychaudhuri equation to study the propagation of sound in these fluids, which, via the Eikonal approximation, correspond to null geodesic congruences in the acoustic spacetimes. We explore this within the acoustic analogues of black holes and cosmological spacetimes. The robustness of the Raychaudhuri equation and the limits of the acoustic analogue are emphasised.

\end{abstract}

\maketitle

\section{Introduction}
There has been a continual interest in the analogue gravity literature to describe fluids in terms of a Lorentzian metric in which an acoustic coupling is explicit \cite{unruh, visser1, visser2}. Within this framework, the propagation of sound is described by by a scalar field $\phi$ obeying the massless minimally coupled scalar field equation

\begin{equation}
    \partial_{\mu}\left(\sqrt{-g} g^{\mu \nu} \partial_{\nu} \phi\right)=0
\end{equation}

Here $g_{\mu \nu}$ is the Lorentzian acoustic metric describing the acoustic spacetime. The line element corresponding to the acoustic metric is
\begin{equation}
    \mathrm{d} s^{2} =\frac{\rho_{0}(\Vec{x},t)}{c_s(\Vec{x},t)}\left[-c_{s}^{2} \mathrm{~d} t^{2}+\delta_{i j}\left(\mathrm{d} x^{i}-v_{0}^{i} \mathrm{~d} t\right)\left(\mathrm{~d} x^{j}-v_{0}^{j} \mathrm{~d} t\right)\right]
\end{equation}

Here, $\rho_0$ is the background fluid density, $c_s$ is the speed of sound and $v^i$ is the fluid velocity field. Imposing additional conditions on these variables will correspond to specific analogue spacetimes.

Further, the Eikonal approximation will be used in describing the scalar field. Within this approximation, we expect the sound waves to travel along null geodesics while the congruences obey the Raychaudhuri equation \cite{ray, lan, ellis} for null case. Written in full, in (3+1) dimensions, the latter can be cast in the form 

\begin{equation}
    \frac{d \Theta}{d \lambda}=-\frac{1}{2} \Theta^2-\sigma^{\mu \nu} \sigma_{\mu \nu}+\omega^{\mu \nu} \omega_{\mu \nu}-R_{\mu \nu} k^\mu k^\nu
\end{equation}

where $\Theta$ is the expansion scalar, $\sigma_{\mu \nu}$ is the shear tensor, $\omega_{\mu \nu}$ is the rotation tensor, and the $R_{\mu \nu}$ is the usual Ricci tensor. Further, $\lambda$ is the affine parameter, and $k^{\mu}$ is the vector field which is tangent to the geodesics. Note that
$\Theta$ can be represented by the divergence form $\Theta=\nabla_\mu k^\mu$.
The acoustic coupling, which is constructed within the confines of non-relativistic fluid mechanics, gives way to analogue models as counterparts of general relativity (GR) spacetimes that embrace the kinematic effects of GR \cite{visser1}. The purpose of this work is to look at the behaviour of the congruence of geodesics in such analogue spacetimes. Another interesting query that deserves a close study is the role of Raychaudhuri equation which is basically concerned with the background geometry of the manifold rather than addressing any specific feature of GR which dictates, in particular, the kinematics of the model involved. \cite{ kar1, kar2, kar3}

The present paper is organized as follows. In section 2, we briefly sketch how the Eikonal approximation. This lets us describe the sound waves using null geodesics. In section 3, we address acoustic black holes, in particular the canonical acoustic black hole. We look at null geodesic congruences, derive the Raychaudhuri equation and show that it is in accordance with what is obtained from the Schwarzschild metric. We briefly discuss other acoustic black hole models. In section 4, we discuss the acoustic analogue of the FLRW cosmology, wherein the curvature of the spacetime is obtained by varying the speed of sound with time. In this context, we explore null geodesic congruences and the Raychaudhuri equation and try to interpret their physical meaning. Finally, a summary of our work is presented. 

\section{The Eikonal Approximation}
Known also as the geometrical optics approximation, the Eikonal approximation is an analytical tool to approximate waves with rays, and hence with geodesics when the space(time) is curved. For a scalar field the approximation gives:
\begin{equation}
    \phi(r, t)=\mathcal{A}(r, t) \exp [\mp i \varphi(r, t)]
\end{equation}

The spatial and temporal dependence of the phase $\varphi(r, t)$  can be further separated, given the geometry is slowly evolving ($\omega \gg \max \{|\dot{c} / c|,|\dot{v} / v|\}$)
\begin{equation}
    \varphi(r, t)=\omega t-\int^r K\left(r^{\prime}\right) \mathrm{d} r^{\prime}
\end{equation}

The condition of restricting the analysis to high values of $\omega$ is at the core of the approximation.

Plugging for $\phi$ into the minimally coupled scalar field equation,
\begin{equation}
    \partial_{\mu}\left(\sqrt{-g} g^{\mu \nu} \partial_{\nu} \phi\right)=0
\end{equation}

The equation corresponding to $\varphi$ turns out to be
\begin{equation}
    g^{\mu \nu} \partial_\mu \varphi \partial_\nu \varphi=0
\end{equation}

Upon making the identification $\partial_\mu \varphi= k_\mu$, we get $k_\mu$ to be a null vector such that
\begin{equation}
    k^\mu k_\mu = 0
\end{equation}

Acting with the covariant derivative on equation(8) and exploiting the fact that it is symmetric $\nabla_\mu k_\nu = \nabla_\nu k_\mu$, we end up with
\begin{equation}
    k^\nu \nabla_\nu k_\mu = 0
\end{equation}
These in turn are the geodesic equations. We see that within the validity of the Eikonal approximation, the waves travel along null geodesics. Choosing an entire non-intersecting family of such geodesics, we can construct null geodesic congruences. These congruences, representing high frequency sound wavefronts, would thus obey the Raychaudhuri equation for the null case, as we shall explore in further sections. 

\section{Acoustic Black holes}

We will begin by looking at acoustic analogues to black hole spacetimes. In this case, the analogy is theoretically well established with the fluid velocity being responsible for the \say{curvature}. Here we look at the two most well known models \textbf{--} the Draining bathtub and the Canonical acoustic spacetimes. The (2+1) dimensional draining bath-tub, which is a rotating solution, is especially relevant for its experimental feasibility. The canonical acoustic solution is stationary and spherically symmetric, and thus offers a close analogue for Schwarszchild-like spacetimes.

\subsection{The (2+1)-dimensional draining bathtub spacetime}

Starting with equation (2) a draining bathtub model in (2+1) dimensions can be constructed that has a sink at the origin.
The conditions imposed on the fluid are consistent with the velocity potential 
\begin{equation}
    \vec{v}=\frac{(A \hat{r}+B \hat{\theta})}{r}
\end{equation}

It leads to the following form of the acoustic metric 
\begin{equation}
    d s^2=- d t^2+\left(d r-\frac{A}{r} d t\right)^2+\left(r d \theta-\frac{B}{r} d t\right)^2
\end{equation}

which can be reset as
\begin{equation}
   d s^2=-\left(1-\frac{A^2+B^2}{r^2}\right) d t^2-2 \frac{A}{r} d r d t-2 B d \theta d t+d r^2+r^2 d \theta^2
\end{equation}

We must note, that we have implicitly assumed cylindrical symmetry and chosen a constant $z$ slice to reduce the number of effective dimensions.
It is worthwhile to explore the behaviour of the null geodesic congruences in such a scenario essentially because it admits both the features of rotation, and a vortex, and is the closest known analogue to Kerr black holes.

Geodesic congruences in the context of Draining bathtub spacetimes of a similar kind, have been extensively discussed in the literature by Dolan \textit{et al} \cite{dolan1, dolan2, dolan3}. 

However, for completeness, we will briefly sketch the form of the null Raychaudhuri equation. Firstly we note that the rotation tensor $\omega_{\mu \nu}$ vanishes. This is because, in the case of outward (or inward) pointing geodesics, the congruences are hypersurface-orthoghonal. This is, in fact, closely related to the definition of the rotation tensor to be symmetric. Further, the shear tensor vanishes identically. This is related to the fact that the shear tensor $\sigma_{\mu \nu}$ is traceless and in (2+1) dimensions the null congruence is codimension one. Therefore, there is no trace-free contribution.

The Raychaudhuri equation in (2+1) dimensions thus reduces to

\begin{equation}
    \frac{d \Theta}{d \lambda}=- \Theta^2 - R_{\mu \nu} k^\mu k^\nu
\end{equation}

Further, the Ricci tensor vanishes. This is to be expected, as we are looking at analogues to spacetimes which are vaccumm solutions to the Einstein equations. We are thus left with

\begin{equation}
    \frac{d \Theta}{d \lambda}=- \Theta^2 
\end{equation}

$\Theta$ upon solving comes out to be $\frac{1}{r}$. The shortcomings of the expansional scalar in this context and the resolution in terms of the Van Vleck determinant are discussed in \cite{thesis}.

\subsection{The Canonical Acoustic Black hole spacetime}
We start with the incompressible acoustic metric (2) which reads
\begin{equation}
    \mathrm{d} s^{2} =\frac{\rho_{0}}{c_s}\left[-c_{s}^{2} \mathrm{~d} t^{2}+\delta_{i j}\left(\mathrm{d} x^{i}-v_{0}^{i} \mathrm{~d} t\right)\left(\mathrm{~d} x^{j}-v_{0}^{j} \mathrm{~d} t\right)\right]
\end{equation}

For this case the speed of sound $c_s$ is taken to a constant independent of time. This leads to the time independence of $\rho_{0}$, as $c_s$ is merely a relation between fluid density and the corresponding pressure. Further, imposing spherical symmetry it implies that
$\rho_{0}$ has to be position independent. This, from the assumption of the barotropic nature of the fluid, points to the position-independence of pressure as well as that of the speed of the sound. Since from the continuity equation it transpires that $v_{0} \propto 1 / r^2$,

We can project, up to a normalisation constant $r_{0}$, the fluid velocity in the manner
\begin{equation}
v_{0}=c_{s} \frac{r_0^2}{r^2}
\end{equation}

Ignoring an irrelevant position dependent factor, we can express the acoustic metric in the form
\begin{equation}
    d s^2=-c_{s}^2 d t^2+\left(d r \pm c_{s} \frac{r_0^2}{r^2} d t\right)^2+r^2\left(d \theta^2+\sin ^2 \theta d \phi^2\right) .
\end{equation}

At this point, we observe that on making the following coordinate transformation
\begin{equation}
    d \tau=d t \mp \frac{r_0^2 / r^2}{c_{s}\left[1-\left(r_0^4 / r^4\right)\right]} d r
\end{equation}

We can map the line element (17) to the canonical structure \cite{visser1}
\begin{equation}
d s^2=-c_{s}^2\left[1-\left(r_0^4 / r^4\right)\right] d \tau^2+\frac{d r^2}{1-\left(r_0^4 / r^4\right)}+r^2\left(d \theta^2+\sin ^2 \theta d \phi^2\right)
\end{equation}

Remark: The $r$ dependence of the above line element is different from that of the Schwarszchild line element. However, in the near horizon approximation $(r \approx r_0)$, the $r$ dependence, upon factorisation can be put in a form proportional to $1/r$.

For the rest of the section we assume, without loss of generality, the convention $c_{s}=1$ and $r_0=1$

We now define a quantity $\beta$ as given by 
\begin{equation}
    \beta(r)=1-\frac{1}{r^4}
\end{equation}

This transforms (19) to 
\begin{equation}
d s^2=-\beta(r) d \tau^2+(\beta(r))^{-1} d r^2+r^2 d \Omega^2 
\end{equation}

For radial null geodesics, $d \theta=d \phi=0$ which means that the rotation ($\omega_{\mu \nu}$) and stress ($\sigma_{\mu \nu}$) tensors do not contribute and the line element is simplified to
\begin{equation}
    d s^2=-\beta(r) d \tau^2+(\beta(r))^{-1} d r^2=0
\end{equation}

Introducing the so-called tortoise coordinate $r_{*}$ satisfying the differential equation 
\begin{equation}
    \frac{d r_{*}}{dr}=\sqrt{-\frac{g_{rr}}{g_{00}}}=\sqrt{\frac{1}{\beta^{2}(r)}}=\frac{1}{\beta(r)}
\end{equation}

and which integrates to 
\begin{equation}
    r_*  =\frac{-\ln (|r+1|)-2 \tan ^{-1}(r)+\ln (|r-1|)}{4}+r
\end{equation}

the line element in terms of $r_*$ becomes
\begin{equation}
    d s^2=\beta(r)\left(-d \tau^2+dr_*^2\right)=0
\end{equation}

In terms of Eddington-Finkelstein coordinates which defines a pair of coordinates $(u, v)$
\begin{equation}
    u=\tau-r_*, \quad  v=\tau+r_*
\end{equation}

we easily notice that $u$ is constant for outgoing null geodesics, while $v$ is constant for incoming null geodesics. Focusing on the outgoing geodesics, we define the corresponding tangent vector field 
\begin{equation}
    k_{\mu}=-\partial_{\mu}u=(-1, (\beta(r))^{-1},0,0)
\end{equation}

such that
\begin{equation}
    k^\mu=g^{\mu \nu} k_\nu=\left((\beta(r))^{-1}, 1,0,0\right)
\end{equation}

The calculations are relevant for the consequences we draw at the end. Hence we will put in several of the steps.
The expansion scalar $\Theta$ can now be calculated by using its representation
\begin{equation}
     \Theta=\nabla_\mu k^\mu=\partial_\mu k^\mu+\Gamma_{\mu_\alpha}^\mu k^\alpha
\end{equation}

where the non-vanishing terms are
\begin{equation}
    \Theta=\Gamma_{01}^0 k^{1}+\Gamma_{11}^{1} k^{1}+\Gamma_{21}^2 k^{1}+\Gamma_{31}^3 k^{1}
\end{equation}

Noting that

$$\Gamma_{01}^0 k^{1}=  \frac{1}{2}\left((\beta(r))^{-1}\right) \frac{d \beta(r)}{d r}$$

$$\Gamma_{11}^{1} k^{1}=-\frac{1}{2}\left((\beta(r))^{-1}\right)\frac{d \beta(r)}{d r}$$

$$\Gamma_{21}^2 k^{1}= \frac{1}{r}$$
$$\Gamma_{31}^3 k^{1}= \frac{1}{r}$$
$\Theta$ is easily estimated to be
\begin{equation}
    \Theta=\frac{2}{r}
\end{equation}

A similar calculation can be performed for incoming geodesics with $k_{\mu}$ defined as 
\begin{equation}
     k_{\mu}=-\partial_{\mu}v
\end{equation}

Then $\Theta$ turns out to be
\begin{equation}
    \Theta=-\frac{2}{r}
\end{equation}

Notice this is twice of what we mentioned for the (2+1) dimensional spacetime. This is because in this case, there is an extra contribution in the sum from the third spatial dimension.

As, in the previous case, $\sigma_{\mu \nu}$, $\omega_{\mu \nu}$ and $R_{\mu \nu}$ all vanish.

Raychaudhuri equation for either case yields
\begin{equation}
   \frac{d \Theta}{d \lambda}=-\frac{2}{r^2}
\end{equation}

Equations (31) and (33) and equation (34) match exactly with the result following from the Schwarzschild metric \cite{poi}. This is because, in the present calculation the explicit form of $\beta(r)$ was not required to be implemented. Note that $\beta(r)=1-\frac{2M}{r}$ would correspond to the Schwarzschild case.
Thus our result for null congruences would hold more generally for any time independent, spherically symmetric metric which satisfies the form prescribed in equation (21). 

As a trivial consequence, we see that the above result holds for $\beta(r)=1$. Within the acoustic framework, this can be obtained by taking the fluid velocity 3-field $v_0=0$ everywhere in equation (2) and further taking $c_s$ to be a constant in time. This is analogous to the Minkowski spacetime. Therefore, in homogenous fluids, the dynamics of high frequency sound waves can be studied using the Raychaudhuri equation. It would be interesting to study non-radial geodesics for this case, which we shall try to explore in a future work.

\section{Acoustic FLRW cosmologies} 
We now turn towards an analogue cosmological model obtained from the acoustic metric. As Barcelo \textit{et al} \cite{anaFLRW} pointed out, the appropriate acoustic analogue of a flat FLRW metric, is obtained by keeping the background fluid at rest $(v_0=0)$ and instead varying the speed of the excitations ($c_s(t)$) within them.

Starting with equation (2), in spherical coordinates two more conditions are imposed.

\begin{enumerate}
    \item The background fluid is considered to be at rest, $v_z=0$. The continuity equation gives that $\rho_0$ is a constant.
    \item Spatially homogenity is assumed. This implies that $c_s$ is independent of $r$.
\end{enumerate}

These lead to
\begin{equation}
   \mathrm{d} s^{2} =\frac{\rho_{0}}{c_s(t)}\left[- c_s^{2}\mathrm{~d} t^{2}+\mathrm{d} \Vec{x}^2 \right]. 
\end{equation}

Further, the metric can be rescaled by a constant factor $\frac{c_0}{\rho_0}$. Here $c_0$ is some convenient reference speed. The line element thus becomes
\begin{equation}
   \mathrm{d} s^{2} =\frac{c_{0}}{c_s(t)}\left[- c_s^{2}(t)\mathrm{~d} t^{2}+\mathrm{d} z^2 \right]=-c_0c_s(t)\mathrm{~d}t^2 + .\frac{c_{0}}{c_s(t)} \mathrm{d} \Vec{x}^2
\end{equation}

In order to convert this line element into the traditional FLRW form, a pseudo time is introduced. This pseudo time $\tau$ is related to the laboratory time $t$ such that 
\begin{equation}
    d \tau = dt \sqrt{c_s(t)/c_0}
\end{equation}

Equation (36) becomes
\begin{equation}
    \mathrm{d} s^{2} = -c_0^2 \mathrm{~d} \tau^{2} + \frac{c_0}{c_s(t)}\mathrm{d} \Vec{x}^2 = -c_0^2 \mathrm{~d} \tau^{2} + \frac{c_0^2}{\Bar{c}_s^2(\tau)}\mathrm{d} \Vec{x}^2
\end{equation}

Here, $\Bar{c}_s(\tau)$ is the speed of sound in terms of the pseudo time. This can be obtained via the chain rule
\begin{equation}
    \Bar{c}_s(\tau)= \frac{d \Vec{x}}{d \tau}= \frac{d t}{d \tau} \frac{d \Vec{x}}{d t} = \sqrt{c_0c_s(t)}
\end{equation}

Equation (38) is completely equivalent to the flat FLRW line element upon making the identification
$$
    c_0d \tau \sim c  dt_{F}  
    $$ 
    and
$$
\frac{c_0^2}{\Bar{c}_s^2(\tau)} \sim a^2(t_{F})
$$

Here $c$ is the speed of light and $t_F$ is the comoving FLRW time. It is worth re-emphasising that within this model the \say{curvature} is obtained by adding time dependence to the speed of sound. This is unlike the case for acoustic black holes where the speed of sound was taken to be constant in time, and the curvature was a consequence of inhomogenous fluid flow. This particular analogue model is also interesting because here we have an analogy with a non-vaccumm solution of GR. For works exploring analogue cosmologies see, \cite{ jain, lang, pandey, vissern}.

For convenience of notation we will choose $c_0=1$.  Having obtained the metric, we would like to construct the various geometric quantities out of it. We shall, again, focus on radial null geodesics. 
For outgoing geodesics, we define the tangent vector field

\begin{equation}
    k_{\mu}= -\partial_{\mu}\left( \Bar{c}_s d\tau - d r  \right) = \left( -\Bar{c}_s(\tau), 1, 0, 0\right)
\end{equation}

Such that 

\begin{equation}
    k^{\mu}= \left(\Bar{c}_s(\tau), \Bar{c}^2_s(\tau), 0, 0  \right)
\end{equation}

This ensures, $k_{\mu}k^{\mu}=0$

Upon solving for the expansion factor, we end up with

\begin{equation}
    \Theta = 2 \Bar{c}_s^2(\tau)\left(-\frac{\Dot{\Bar{c}}_s(\tau)}{\Bar{c}_s(\tau)}  + \frac{1}{r}   \right)
\end{equation}

The dot represents a derivative with respect to the Pseudo time $\tau$. Note that equation (39) tells us that this pseudo time is linearly related to the laboratory time $t$. This is important as in this case, the qualitative features of the dynamics will get carried over, even when using $t$. 

There are a couple of things to notice about the expression for $\Theta$. Firstly, for $\Dot{\Bar{c}}_s(\tau)=0$, the expression reduces to

$$
     \Theta =  \Bar{c}_s^2(\tau)\frac{2}{r} 
$$

This is upto a constant, the same as equation (26). This was expected  as $\Dot{\Bar{c}}_s(\tau)=0$ would correspond to a static (and flat) acoustic spacetime. Further, the minus sign in the $\Dot{\Bar{c}}_s(\tau)$ term in (44) ensures that in order to get an expanding (analogue) spacetime, a decreasing speed of sound is required. 

Now, we move on to calculate the Ricci tensor. The relevant non-vanishing terms are
\begin{equation}
    R_{00}= 3 \left(\frac{\Ddot{\Bar{c}}_s}{c_s} - \frac{1}{2} \frac{\Dot{\Bar{c}}^2_s}{\Bar{c}^2_s}   \right)
\end{equation}
and

\begin{equation}
    R_{11}= \frac{5}{4}\frac{\Dot{\Bar{c}}^2_s}{\Bar{c}^4_s} - \frac{\Ddot{\Bar{c}}_s}{c^3_s}
\end{equation}

Here, $\Bar{c}_s(\tau)$ is written as $\Bar{c}_s$ to avoid cluttering the equations.

Spatial homogeneity and isotropy together with the fact that we are looking at hypersurface orthogonal geodesics ensure that $\sigma_{\mu \nu}$ and $\omega_{\mu \nu}$ will vanish.

Writing out the Raychaudhuri equation in full gives us 

\begin{equation}
   \begin{aligned}
    \frac{d \Theta}{d \lambda} &= - \frac{1}{2}\Theta^2 - R_{\mu \nu}k^{\mu}k^{\nu} \\
    &= \Ddot{\Bar{c}}_s\left( \frac{1}{\Bar{c}_s}-3\Bar{c}_s \right) + \Dot{\Bar{c}}_s \left( \frac{7}{4}\Dot{\Bar{c}}_s - 2 c^2_s \Dot{\Bar{c}}_s + \frac{4}{r\Bar{c}^3_s}\right) + \frac{2\Bar{c}^4_s}{r^2}
    \end{aligned}
\end{equation}

This equation puts forward the various competing terms generating the dynamics of the expansion scalar. As expected there is a  contribution from the second derivative of the scale factor (inverse of $c_s$). It could be interesting to think about this equation in a fluid mechanical context. It might give alternate geometry based explanations to known results. 

However, this equation does not quite resemble the form for the null Raychaudhuri equation as is usually seen in cosmology textbooks. This is so because, in writing down that particular form of the Raychaudhuri equation, the Freidmann equation is implicitly used. Our inability to do that comes from the fact that the acoustic analogue of gravity can mimic only the kinematic effects of GR. That is, within the acoustic framework the dynamics of GR, namely, Einstein equations and consequently the (first) Freidmann equation cannot be reproduced. Hence we do not expect here to obtain the form involving terms from the energy-momentum tensor.

\section{Summary}
In this work, we emphasised the robustness of the Raychaudhuri equation by pointing out its relevance in an analogue spacetime situation where we did not have to make any use of Einstein equations or the null energy condition \cite{visser4, wald}. We noted the similarities and differences in the results we obtained here with those in GR. Therefore, we pointed out the limits of the acoustic analogue for gravity, which is well known to mimic only the kinematic effects of GR\cite{visser5}. Finally, we provided an illustrative example of how the mathematical methods constructed for GR could be used to understand the behaviour of sound in fluids. 

We would also like to mention a query which we believe deserves some investigation. It is the following \textbf{---} Upon quantising the fluid excitations, we end up with phonons. Analysing the validity and consequences of the Quantum Raychaudhuri Equation (QRE) \cite{das} in this framework could be valuable for pracitioners of both analogue gravity and phonon-based physics. It could possibly be of interest to physicists studying the foundations of quantum mechanics as the QRE is set within the Bohmian \cite{bohm1, bohm2} formulation of Quantum Mechanics. 

\section*{Acknowledgements}
I wish to thank Prof. Bijan Bagchi for the insightful discussions and for his comments on the early draft of this manuscript. I also wish to thank Rahul Ghosh and Sauvik Sen for their valuable feedback.


\begin{thebibliography}{20}
\small
\bibitem{unruh}  Unruh W. G. 1981 \textit{Phys. Rev. Lett.} \textbf{46}, 1351.

\bibitem{visser1}  Visser M. 1998  \textit{Class. Quant. Grav.} \textbf{15}, 1767.
\bibitem{visser2}  Barceló C., Liberati S., and Visser M. 2011 \textit{Living Rev. Relativ.} \textbf{14} 1-159.

\bibitem{ray} Raychaudhuri A. 1955 \textit{ Phys. Rev. }\textbf{98}, 1123.
\bibitem{lan} Lifshitz E.M. and Khalatnikov I. M. 1963 \textit{Adv. Phys.} \textbf{12}, 185.
\bibitem{ellis} Ellis G. F. 2007 \textit{ Pramana}, \textbf{69}.
%Acoustic black holes,


\bibitem{poi}  Poisson E. 2004 A Relativist's Toolkit, 250.
\bibitem{kar1} Kar S. 2001 \textit{Phys. Rev.} \textbf{D64}, 105017.
\bibitem{kar2}Kar S. and SenGupta S. 2007
\textit{Pramana} \textbf{69}, 49.
\bibitem{kar3} Kar S. 1996 \textit{Phys. Rev.}  \textbf{D54}, 6408.
\bibitem{dolan1} Torres, T., Lloyd M., Dolan S.R. and Weinfurtner S. 2022  \textit{Phys. Rev. Research} \textbf{4(3)}.
\bibitem{dolan2}Dempsey D. and Dolan S.R. 2016 \textit{Int. J. Mod. Phys.} \textbf{D 25} (09).

\bibitem{dolan3} Dolan S.R., Oliveira L.A. and Crispino L.C. 2012  \textit{Phys. Rev.} \textbf{D 85} (4).


\bibitem{thesis} Dempsey D. 2017 \textit{Wave propagation on black hole spacetimes}.

\bibitem{anaFLRW} Barcel\'{o} C., Liberati, S. and Visser M. 2003 \textit{Int. J. Mod. Phys.} \textbf{D12}, 09.

\bibitem{jain} Jain P,  Weinfurtner S., Visser M. and  Gardiner C. W. 2007 \textit{Phys. Rev.}
\textbf{A 76}, 033616. 
\bibitem{pandey} Pandey A. 2023 \say{A note on Analogue semi-classical gravity in (1+1) dimensions.} (unpublished).
\bibitem{lang}  Lang S. and  Sch\"{u}tzhold R. 2019 \textit{Phys. Rev.} \textbf{D 100}, 065003.
\bibitem{vissern}  Barcel\'{o} C., Liberati S. and  Visser M. 2003 \textit{Phys. Rev.} \textbf{A 68}, 053613.

\bibitem{visser4} Visser M. 1996 \say{Lorentzian Wormholes. From Einstein to Hawking.}
\bibitem{wald} Wald R. 1984 \say{General Relativity.}
\bibitem{visser5}Visser M. 2003 \textit{Int. J. Mod. Phys.}, \textbf{D12}.

\bibitem{das}Das S. 2014 \textit{Physical Review} \textbf{D89} 8.
\bibitem{bohm1} Bohm D. 1952 \textit{Phys. Rev.} \textbf{85} 166.
\bibitem{bohm2} Bohm D., Hiley B. J. and Kaloyerou P. N. 1987 \textit{Phys. Rep.} \textbf{144} 321.






\end{thebibliography}
\end{document}